\documentclass[useAMS,usenatbib,usegraphicx,twocolumn,usedcolumn]{mn2e}
 \usepackage[T1]{fontenc}
 \usepackage{aecompl}
\usepackage{float}
\usepackage{graphicx}
\usepackage{grffile}
\usepackage{lineno}
\usepackage{multirow}
\usepackage{subfigure}
\usepackage{color}
\usepackage{cleveref}
\def\vFv{\nu F_{\nu}}
\def\Ep{E_{\rm p}}

\title[On Spectral Evolution and Temporal Binning in Gamma-Ray
Bursts]{On Spectral Evolution and Temporal Binning in Gamma-Ray
  Bursts} \author[J. Michael Burgess]{J. Michael
  Burgess$^{1,2}$\thanks{E-mail:
    jamesb@kth.se (JMB)}\\
  $^{1}$The Oskar Klein Centre for Cosmoparticle Physics,
  AlbaNova, SE-106 91 Stockholm, Sweden\\
  $^{2}$Department of Physics, KTH Royal Institute of Technology, AlbaNova, SE-106 91 Stockholm, Sweden\\
}
\begin{document}

\date{Accepted XXXX December XX. Received XXXX December XX; in original form XXXX October XX}

\pagerange{\pageref{firstpage}--\pageref{lastpage}} \pubyear{2014}

\maketitle

\label{firstpage}

\begin{abstract}
  The understanding of the prompt $\gamma$-ray spectra of gamma-ray
  bursts (GRBs) is of great importance to correctly interpreting the
  physical mechanisms that produce the underlying event as well as the
  structure of the relativistic jet from which the emission
  emanates. Time-resolved analysis of these spectra is the main method
  of extracting information from the data. In this work, several
  techniques for temporal binning of GRB spectra are examined to
  understand the systematics associated with each with the goal of
  finding the best method(s) to bin lightcurves for analysis. The
  following binning methods are examined: constant cadence (CC),
  Bayesian blocks (BB), signal-to-noise (S/N) and Knuth bins (KB). I
  find that both the KB and BB methods reconstruct the intrinsic
  spectral evolution accurately while the S/N method fails in most
  cases. The CC method is accurate when the cadence is not too coarse
  but does not necessarily bin the data based on the true source
  variability. Additionally, the integrated pulse properties are
  investigated and compared to the time-resolved properties. If
  intrinsic spectral evolution is present then the integrated
  properties are not useful in identifying physical and cosmological
  properties of GRBs without knowing the physical emission mechanism
  and its evolution.

\end{abstract}


\section{Introduction}
Spectral evolution has long been studied in the context of GRBs
\citep{1983Natur.306..451G,Liang:1996cl,Crider:1998uf,1986ApJ...301..213N,Burgess:2014db,2010ApJ...725..225G}. While
the shape of the spectrum can aid in identifying the type of emission
mechanism occurring in the jet, spectral evolution can elucidate the
temporal evolution of the density, magnetic field, and structure of
the jet. The unprecedented spectral and temporal resolution of the
{\it Fermi} Gamma-ray Burst Monitor (GBM) \citep{2009ApJ...702..791M}
allows for detailed observations of the evolution of GRB spectra. It
is therefore important to evaluate the ability of GBM to measure the
intrinsic spectral evolution of a GRB.

To reconstruct the spectral evolution of GRBs in the data, the
observed lightcurve must be binned in time. Due to the way photons are
detected by GBM, their true energy is not known and the shape of the
spectrum can only be ascertained after the data have been binned in
time and folded through the instrument's detector response
  matrices (DRMs). Therefore, it is impossible to bin the data in
time based on the intrinsic spectral evolution a priori. Binning must
then be based upon a balance between having enough signal to
accurately fit the spectrum and having a fine enough time-resolution
to detected the intrinsic changes in the spectrum over time. Hence,
several methods to bin the data in GRB spectral analysis have been
developed. In this work, these methods are investigated to evaluate
their ability to accurately reconstruct the intrinsic spectral
evolution in GRBs.

Since it is impossible to know the intrinsic spectral evolution in a
GRB, this investigation requires a set of simulated GRBs with known
intrinsic spectral evolution. This is achieved via a simulation code
that can map an evolving photon model into GBM time-tagged event (TTE)
data which can then be analyzed like real source data. These
simulations also afford the ability to examine the integrated
properties of GRB spectra as compared to the known evolution of the
spectra. With the integrated properties being important to using GRBs
as cosmological tools, this investigation can provide insights to the
power of these properties to be indicative of the physical properties
of the source.

This article is organized in the following way: Section \ref{sec:sim}
provides a description of the simulation method used to construct the
control set of GRB pulses, Section \ref{sec:bin} introduces the binning
methods to be investigated, Section \ref{sec:data} describes the simulated
data set that will be binned and analyzed, and Section \ref{sec:results}
investigates the results of the study for both spectral evolution and
the integrated properties of the simulated GRBs.

\section[]{GBM TTE Simulator}
\label{sec:sim}
In order to assess the ability of the various binning methods to
accurately reconstruct the spectral evolution of GRB emission, it is
imperative to create a control set of simulated GRB data with known
spectral evolution. The simulated data must meet two requirements to
be of use in this study:
\begin{itemize}
\item the freedom to rebin the data to arbitrary binnings
\item and data must be folded through GBM DRMs to mimic the response
  of the instrument.
\end{itemize}

Achieving these requirements is not possible with the two widely used
analysis software packages for GRB spectral analysis; XSPEC
\cite{xspec} and
RMFIT\footnote{http://fermi.gsfc.nasa.gov/ssc/data/analysis/rmfit/}. Therefore,
a simulator was designed that can generate GBM TTE data for any photon
model that is a function of time and energy.

First, a spectral shape such as the Band function \citep{Band:1993wc}
is chosen as the primary shape to be simulated. The spectral
parameters are given as a function of time yielding a function
$\mathcal{F}_{\rm evo}(\varepsilon,t_{\rm a})$ [phts s$^{-1}$
cm$^{-2}$] that is the equation for the simulated lightcurve. Here,
$\varepsilon$ is the photon energy and $t_{\rm a}$ is the arrival time
of the photon. This equation is then numerically integrated over the
duration of the emission to obtain the bolometric lightcurve so that
the number of photons to be generated is known. From the bolometric
lightcurve, time-tags for the arrival time of photons are generated
using a non-homogeneous Poisson generator. This method randomly
selects arrival times via an inverted Poisson distribution and then
thins the number of time-tags to the shape of the bolometric
lightcurve via an acceptance-rejection sampler. The time-tags will
ultimately be the TTE time-tags in the generated data file.

Once the $t_{\rm a}$'s of all photons are generated, they are input
into $\mathcal{F}_{\rm evo}$ projecting the function into being the
energy distribution of photons at $t=t_{\rm a}$. This energy
distribution is treated as a probability distribution from which
photon energies are randomly selected via a second
acceptance-rejection sampler. After all photons have a time and energy
tag associated with them, the energies must be converted to GBM
pulse-height analysis (PHA) channel. This is performed by selecting a
GBM DRM for each detector that will have a simulated data set. The two
types of detectors on GBM are Sodium Iodide (NaI) for the energy range
8-2000 keV and Bismuth Germanate (BGO) for the energy range 300-40000
keV. The DRMs map photon energy into PHA channel and encode the
physical interactions that occur during photon detection inside the
crystals. Each photon energy row of the response matrix is converted
to a probability distribution in PHA channel space via dividing by the
geometric area of the detector. The geometric area is calculated by
computing the projected area of the detector at the proper incident
angle for the simulated source. This method properly takes into
account the possibility that an incident photon can deposit more than
one count in the detector as a result of multiple scattering.

A homogeneous Poisson background is superimposed on the source data
using the same technique as described above except that the photon
spectrum of the background is assumed to be a simple power law. The
index of the power law used for all simulations in this work is -1.4
and was chosen by fitting the low-energy portion of several GBM
background intervals. However, it was checked that changing the value
of this photon index did not affect the fit results of the source
spectrum. This indicates that the background subtraction technique
used in RMFIT is correctly removing the background from the source
when the background is fit properly.

Once the detector counts are all tagged in channel number and time,
they are saved in the standard GBM TTE FITS file format and can be
analyzed with RMFIT as if they were real data. The TTE
data can then be freely rebinned in time to test various binning
methods.

\section[]{Binning Methods}
\label{sec:bin}
In this section, the various methods for creating time bins for
spectral analysis of GRBs are described. The time evolution of
spectral parameters is key to unraveling the complex emission
mechanisms and jet dynamics in GRBs. Yet, there is no standard method
for binning the data and one is typically selected based upon the
desired purposes of the experiment. Ideally, the method chosen should
be objective. All these methods share a common drawback in that they
cannot bin the data in time based on the spectral evolution of the
GRB. Attempts have been made to do this \citep{Guiriec:2013hl} but
will not be investigated here.

\subsection[]{Constant Cadence}
The simplest method for binning the data is by choosing a constant
cadence (CC) where the time bins are uniform throughout the duration
of the GRB. The method is objective in how the variability of the
burst is treated. For example, once a bin width ($\Delta T$) and start
time ($T_0$) are selected, the choice of bins is completely
determined. Moreover, the choice of binning does not depend on the
flux history or energy distribution of the burst. For this work, three
cadences are selected to bin the data: $\Delta T=5.0s,1.0s,0.5s$
denoted as CC$_5$, CC$_1$, CC$_{0.5}$ respectively.

Drawbacks of this method are that the choice of one bin width for the
duration of the burst, while objective, neglects the fact that the
flux history and spectral shape may change slower or faster than the
chosen cadence.

\subsection[]{Bayesian Blocks}
Bayesian blocks (BBs) \citep{2013ApJ...764..167S} are time bins chosen
such that each bin is consistent with a constant Poisson rate. This is
done by algorithmically subdividing the flux history of the GRB
lightcurve and comparing the likelihood of the distribution of the
count rate of each bin to being piecewise constant or constant. Time
bins selected in this way will have a variable width and variable
signal-to-noise ratio. The selection of the bins will reflect the true
variability of the data which is advantageous for studying changes in
flux.

The method does not insure that there is adequate signal in the bins
to make an accurate spectral fit. This is because it considers
fluctuations in the background on equal footing with the source. It is
therefore beneficial that the data have a somewhat constant background
or that the source be much more intense than the background.

\subsection[]{Knuth Bins}
Similar to BBs, the Knuth binning (KB) method
\citep{2006physics...5197K} seeks to find the optimal binning based on
the data alone using the assumption that the data are best described
by a piecewise constant model. The method of KB differs by removing
the assumption that prior information is known about the intrinsic
density distribution of the bins and instead uses a Bayesian method
that assumes little to no information about the prior distribution and
that the bins all have equal width. The method seeks to find the
simplest model that describes the variability of the data. In this
sense, the method will not find the short duration features found by
BBs but will have more counts or a greater signal-to-noise ratio over
the duration of the burst.

\subsection[]{Signal to Noise}
When performing spectral analysis, it is necessary to have enough
counts above background in the data to fit the spectrum with high
significance. A way of guaranteeing that the ratio of signal counts to
background counts remains constant is by defining bins with a given
signal to noise ratio. To achieve this a background must be selected
in the data. Starting from $T_0$, source and background are
accumulated until the desired ratio is achieved. Herein, the signal to
noise ratio is chosen to be 50 which is a slightly higher than what is
used in the GBM spectral catalogs \cite{Goldstein:2012go}. This
results in varied bin width. Bins having high signal counts are narrow
and those with a low signal flux are wide. While the bins will have
uniform signal-to-noise ratios, the method completely neglects
spectral evolution and the intrinsic flux history of the GRB. This
neglect occurs in the sense that the method does not look back as it
marches forward and bins the data and therefore arbitraily combines
bins with intrinsically different Poisson rates. Therefore, the method
has change points that are a function of both the flux and the chosen
$T_0$.


\section[]{Simulation Data Set}
\label{sec:data}
In order to examine the effects of various binning methods on
reconstructing the intrinsic spectral evolution of GRBs, a set of
simulated GRB data was created. This set consists of single pulses
which can be viewed as the building blocks of more complex
lightcurves.  The spectrum chosen as the basis for the simulations is
the common Band function; ubiquitous in the spectral analysis of
GRBs. The Band function is a smoothly broken power law with a fixed
curvature. It is parameterized by its low and high-energy spectral
indices, $\alpha$ and $\beta$ respectively, as well as its $\vFv$ peak
energy, $\Ep$. The temporal evolution of the Band function's
parameters have been studied in great detail in search for their
physical connection to the outflow. Herein, the evolution of $\Ep$ is
the focus of the investigation. One of the most common found
evolutionary patterns of $\Ep$ is the so-called hard-to-soft evolution
where $\Ep$ evolves from high to low energies over the duration of the
GRB \citep{Band:1997}. Due to its commonality, hard-to-soft evolution
will be the chosen form for these simulations and approximated as a
power law in time:
\begin{equation}
  \label{eq:ep}
  \Ep\left( t \right)\;=\;E_0(t+1)^{-\gamma}.
\end{equation}
Here, $\gamma$ is the decay index of $\Ep$ and four values are
simulated: 1, 1.5, 2, and 2.5. For this study, $E_0$ is chosen to be
2~MeV.

\begin{figure}
\includegraphics[scale=1]{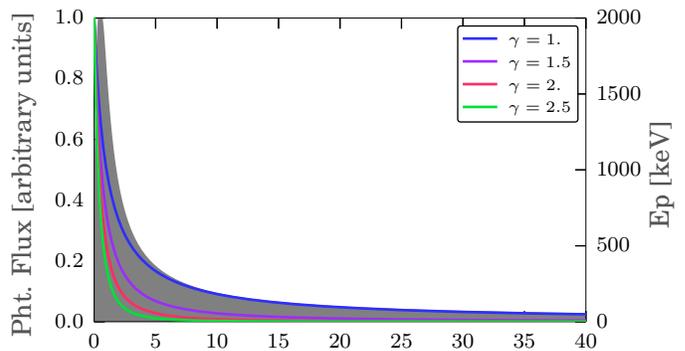}  
 \caption{The intrinsic properties of the simulated datasets. The
   pulse profile is shown in {\it grey} and the different evolutions
   of $\Ep$ are super-imposed on the flux history.}
\label{fig:exEpevo}
\end{figure}

\begin{figure*}
\centering
\subfigure[]{\includegraphics[scale=1]{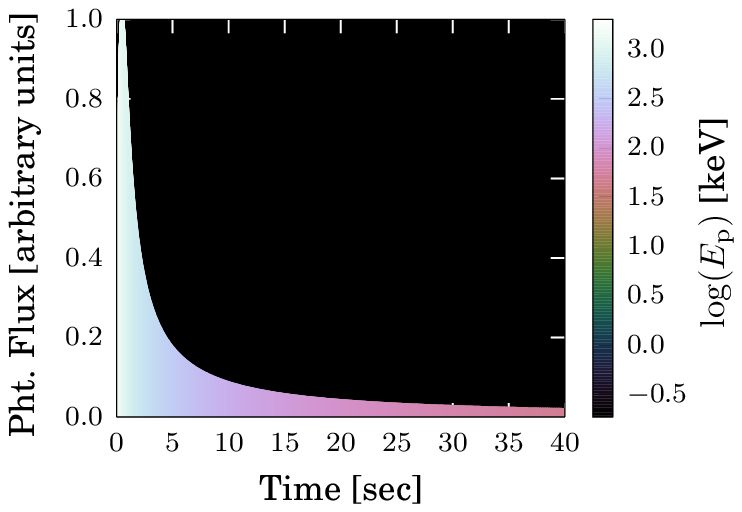}}\subfigure[]{\includegraphics[scale=1]{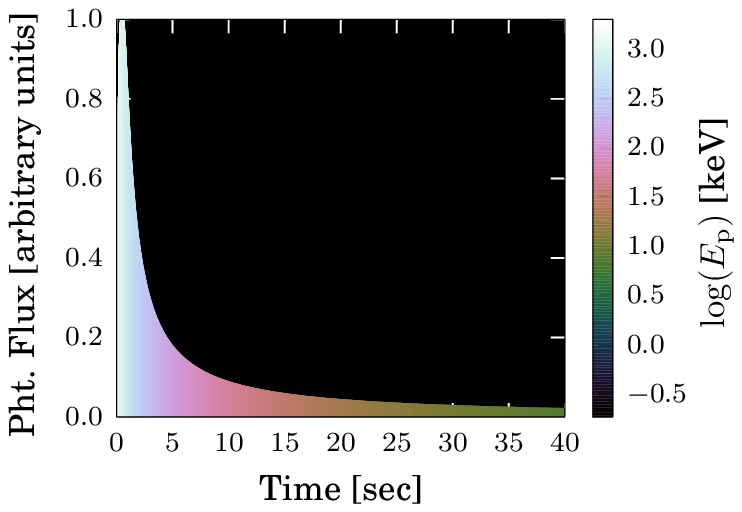}}
\subfigure[]{\includegraphics[scale=1]{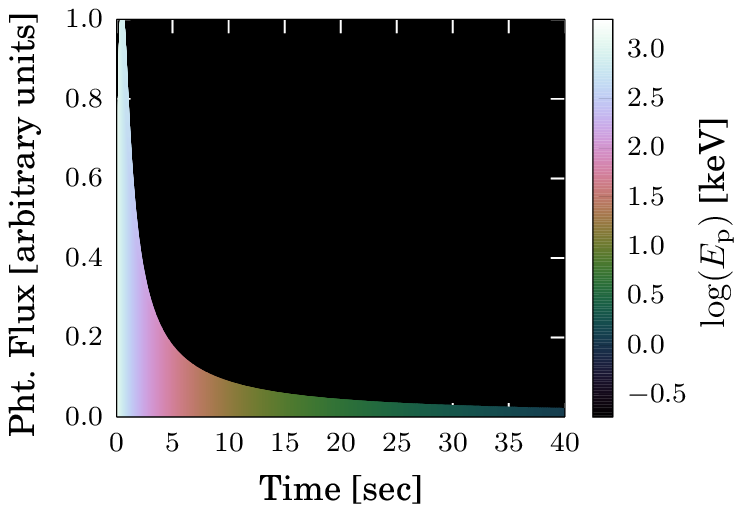}}\subfigure[]{\includegraphics[scale=1]{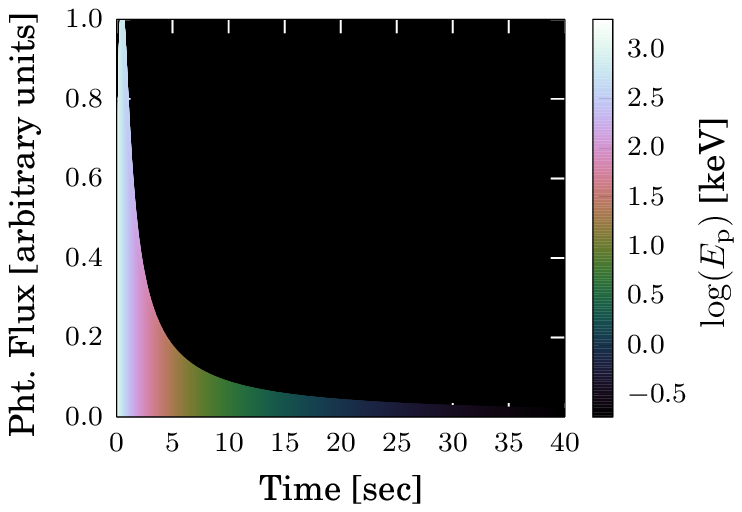}}
\caption{Illustrating the evolution of $\Ep$ as a function of the flux
  history for each choice of $\gamma=$1(a), 1.5(b). 2(c), 2.5(d).}
\label{fig:exEpevo2}
\end{figure*}

To approximate a typical GRB, the amplitude of the Band function is
evolved using the so-called KRL pulse shape given by
\citet{2003ApJ...596..389K}. Since the photon flux is modified by the
value of $\Ep$ and the pulse shape should be the same for different
values of $\gamma$, the amplitude is renormalized so that only the
value the KRL pulse shape determines the flux. This insures that all
simulated datasets have the same photon flux which is important when
calculating the S/N and BB binnings. This introduces an artificial
hardness-intensity correlation but does not affect the investigation
here. The value of $\beta$ is held fixed at its typically observed
value of -2.2 for all simulated data. However, for each value of
$\gamma$, a fixed value of $\alpha$ is chosen from set -1, 0, and
1. Therefore, there are 12 sets of simulated data in total.  With all
the parameters specified, the function $\mathcal{F}_{\rm evo}$ is
completely defined and appears as shown in
Figures \ref{fig:exEpevo} and \ref{fig:exEpevo2}.

\begin{figure}
  \includegraphics[scale=1]{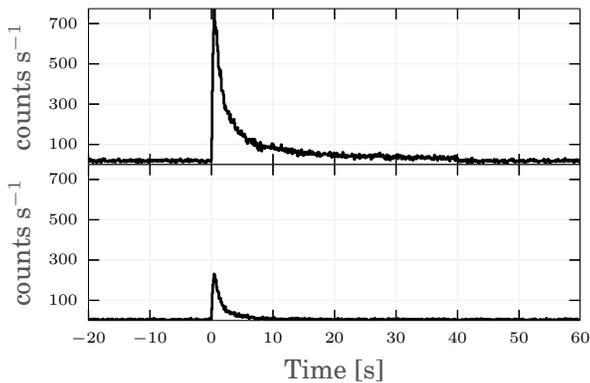}
  \caption{Example NaI ({\it top}) and BGO ({\it bottom}) TTE
    lightcurves generated from $\mathcal{F}_{\rm evo}$. The parameters
    used to generate this simulation are $\gamma=1.$ and
    $\alpha=-1$. The background is a 0-order polynomial in time and is
    distributed in photon number as a power law with spectral index
    -1.4.}
\label{fig:data}
\end{figure}

For each combination of $\gamma$ and $\alpha$, three TTE files are
generated (see Figure \ref{fig:data}); two NaI files and one BGO file. The
GRB pulses are all given a duration of 40 seconds with a super-imposed
background of 80 seconds. The DRMs used to make the TTE files come
from the detection of GRB 110721A \citep{Axelsson:2012ic} and were
chosen such that the source angles to the detector normal are all less
than $60^{\circ}$.

\section[]{Investigating Spectral Evolution}\label{sec:results}
\subsection[]{Analysis Method}
\begin{figure*}
  \centering
  \includegraphics[scale=1.]{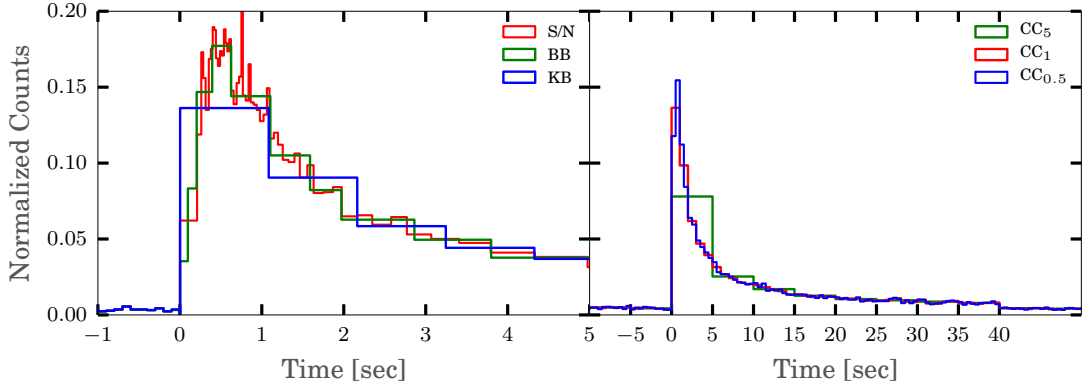}
  \caption{The temporal binning of the lightcurves with the above
    listed methods. The lightcurves have been normalized so they can
    be directly compared. The KB method does not resolve the shape of
    the peak as well as the BB method.}
  \label{fig:binning}
\end{figure*}

Each data set is temporally binned via the methods described in
Section \ref{sec:bin} (see Figure \ref{fig:binning}). For both the BB
and KB methods the data were binned with the routines of the AstroML
software library \citep{VanderPlas:gg}.The backgrounds in each are fit
with a constant background. The signal region of the pulse is selected
and each time bin is fit with the Band function. Near the tail of the
GRB, the weak flux causes some fits to fail and those time bins are
excluded from the study. The photon flux and energy flux $F_{\nu}$ for
each fit is calculated by integrating the model over the full
bandpass.

\subsection{Integrated Pulse Properties}

\begin{figure}
  \includegraphics[scale=1]{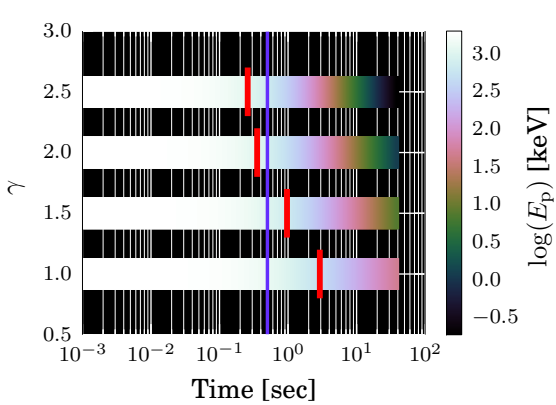}
  \caption{The fitted of $\Ep$ found by fitting the integrated
    spectrum ({\it red}) is super-imposed on the simulated $\Ep$
    evolution for each value of $\gamma$. The {\it purple} line
    indicates the time of peak flux of the simulated pulse. It is
    clear that the integrated $\Ep$ is correlated with the speed of
    the intrinsic spectral evolution but uncorrelated with the peak
    flux.}
  \label{fig:epInt}
\end{figure}

The integrated $\Ep$ is a commonly investigated property of GRBs. Its
correlation with the total $F_{\nu}$ has been used to relate GRBs to
cosmological properties
\citep{Ghirlanda:2004tq,Amati:2003wu,Ghirlanda:2005uy}. It is
important to understand the relevance of the integrated $\Ep$ to its
time evolution with a GRB. In Figure \ref{fig:epInt}, the integrated
$\Ep$ is plotted against its evolution. Clearly, the integrated $\Ep$
is a function of the spectral evolution and has a value that is close
to the maximum simulated $\Ep$. It is not correlated with the value of
$\Ep$ at the time of peak flux. Additionally, the mean fitted $\Ep$ is
calculated from the CC$_{0.5}$ bins and compared to the integrated
$\Ep$ which is systematically higher (see Table \ref{tab:epInt}).

The integrated $F_{\nu}$ is calculated from the integrated fit and
compared to the summed $F_{\nu}$ of the
CC$_{0.5}$. 
The integrated $F_{\nu}$ is systematically less that the summed
$F_{\nu}$. It is therefore difficult to make a connection between the
time-resolved and time-integrated properties.

\begin{table*}
\centering
\begin{tabular}{cccccc}

  $\alpha$ & Index & Integrated $E_{\rm p}$ [keV] & Average $E_{\rm p}$ [keV]& Integrated $F_{\nu}$ [erg s$^{-1}$ cm$^{-2}$]& Summed $F_{\nu}$ [erg s$^{-1}$ cm$^{-2}$]\\
  \hline\hline
  \multirow{4}{*}{-1} & 1.0 & 511.71  & 293.92 & 1.84e-06 & 1.38e-04 \\
           & 1.5 & 725.27  & 369.33 & 1.33e-06 & 9.98e-05 \\
           & 2.0 & 1091.00 & 436.71 & 1.13e-06 & 8.10e-05 \\
           & 2.5 & 1127.36 & 487.78 & 9.50e-07 & 6.61e-05 \\
  \hline
  \multirow{4}{*}{0}  & 1.0 & 867.36  & 241.88 & 4.17e-06 & 3.34e-04 \\
           & 1.5 & 1127.91 & 347.39 & 2.90e-06 & 2.31e-04 \\
           & 2.0 & 1349.98 & 405.60 & 2.23e-06 & 1.73e-04 \\
           & 2.5 & 1296.32 & 482.99 & 1.88e-06 & 1.39e-04 \\

\end{tabular}
\caption{The integrated pulse properties of the simulated data set compared with the time-resolved properties.}
\label{tab:epInt}
\end{table*}

\subsection{Reconstructing the Evolution of $\Ep$}
\begin{figure*}
\centering
\subfigure[]{\includegraphics[scale=1]{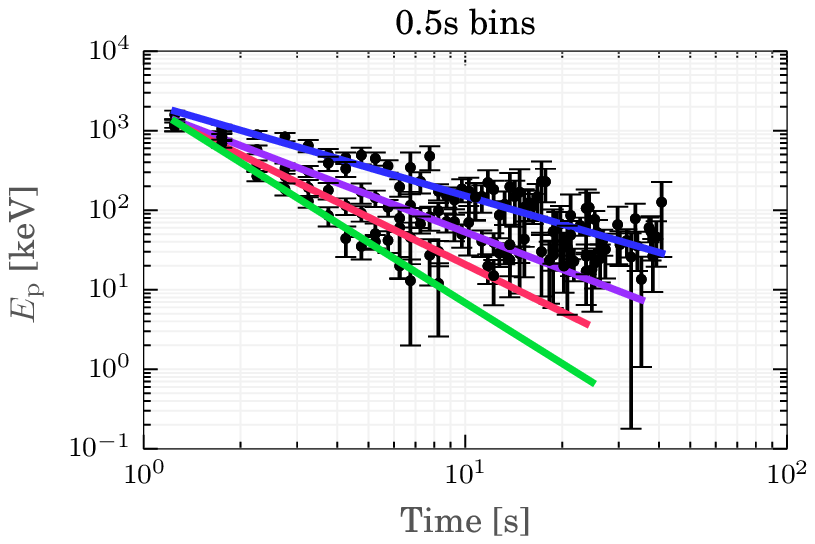}}\subfigure[]{\includegraphics[scale=1]{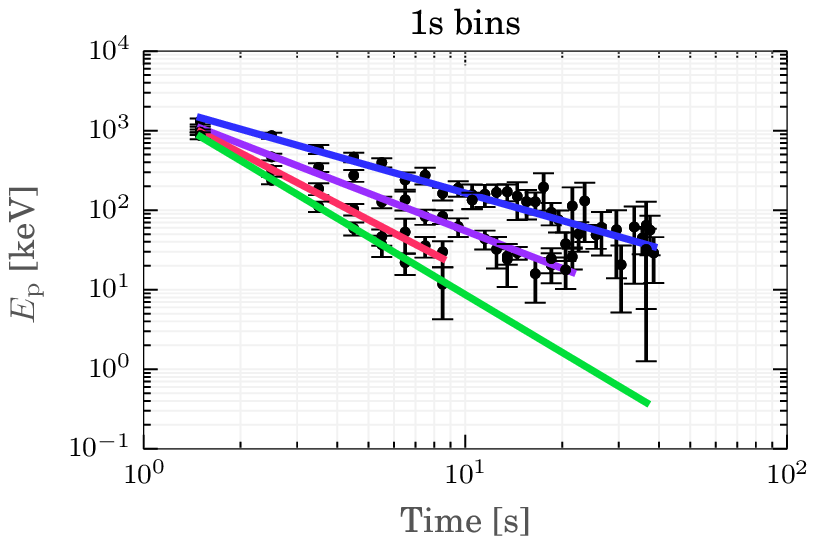}}
\subfigure[]{\includegraphics[scale=1]{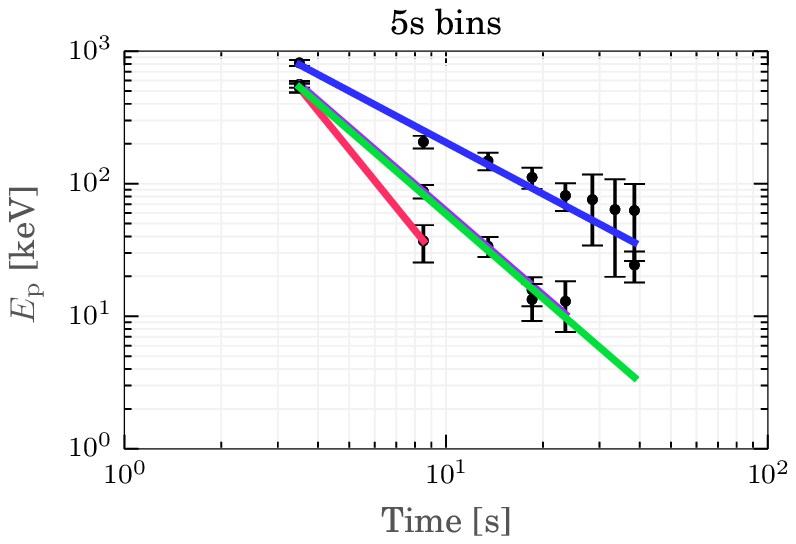}}\subfigure[]{\includegraphics[scale=1]{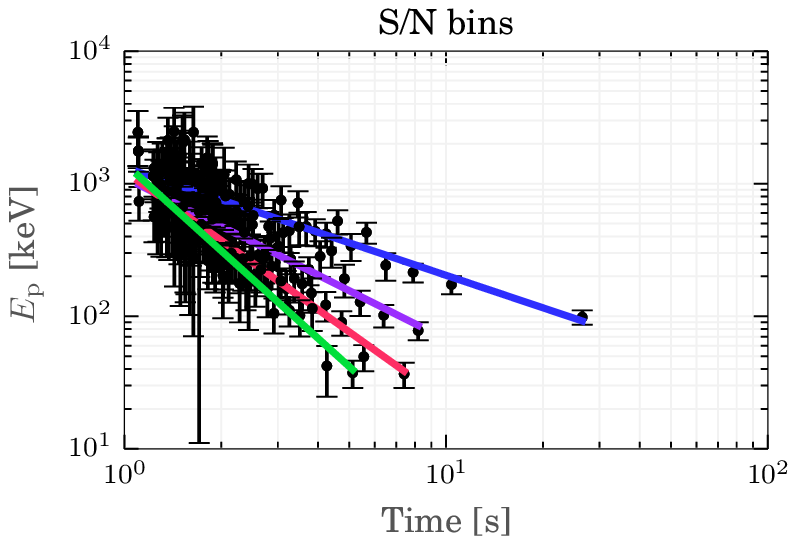}}
\subfigure[]{\includegraphics[scale=1]{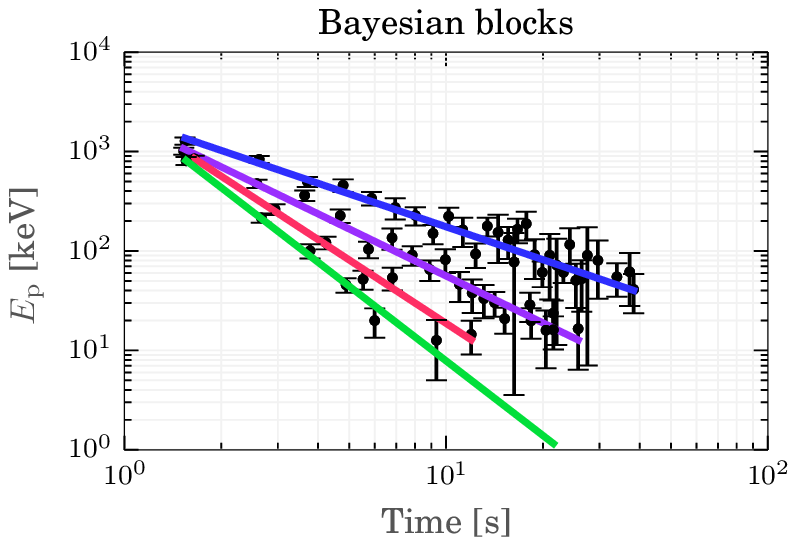}}\subfigure[]{\includegraphics[scale=1]{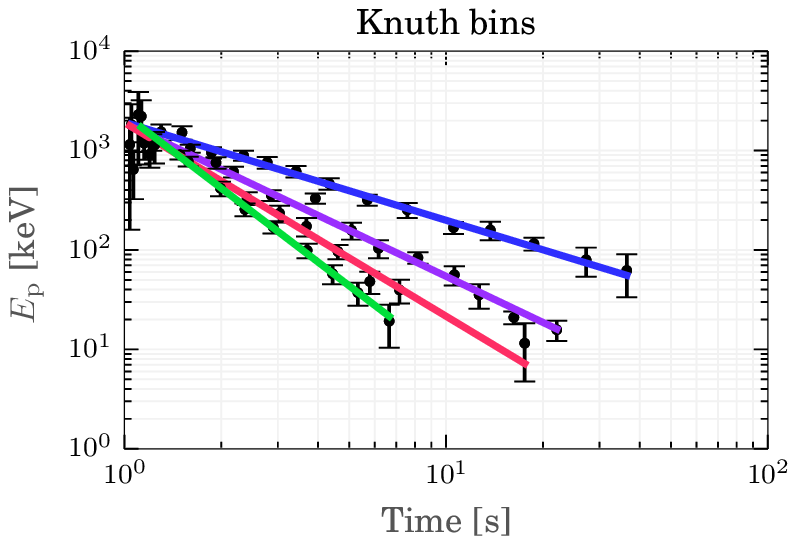}}
\caption{The reconstructed evolution of $\Ep$ for each of the binning methods with simulated $\alpha=-1$. The various fit lines indicate the reconstructed evolution for $\gamma=$1 ({\it blue}), 1.5 ({\it purple}), 2 ({\it pink}), and 2.5 ({\it green}). Some data points are missing due to the fitting engine failing to converge.}
\label{fig:alph-1EpEvo}
\end{figure*}

\begin{figure*}
\centering
\subfigure[]{\includegraphics[scale=1]{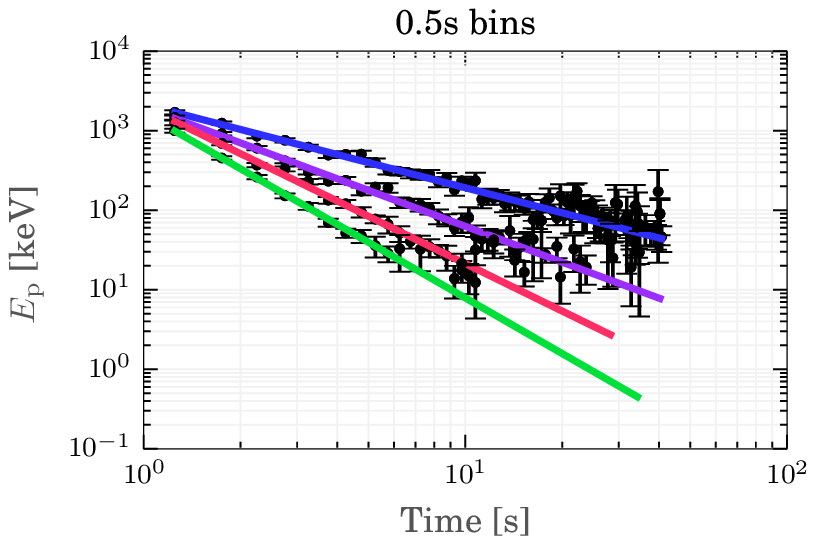}}\subfigure[]{\includegraphics[scale=1]{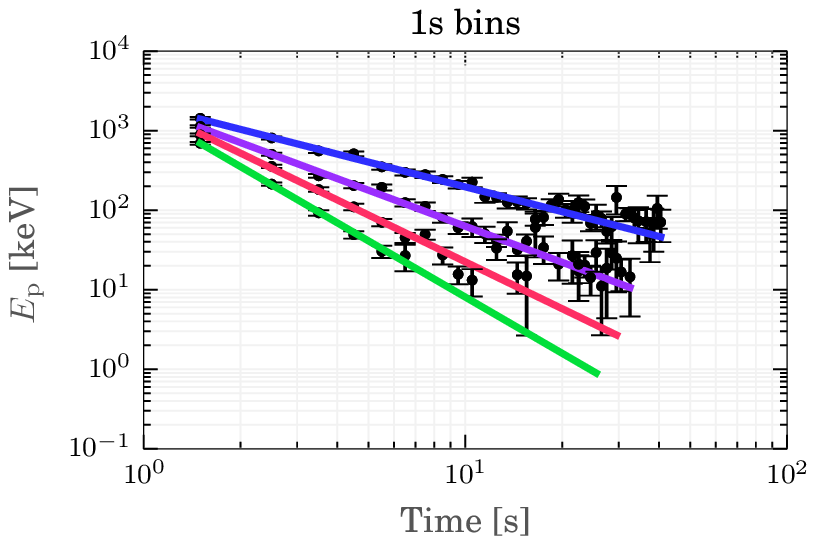}}
\subfigure[]{\includegraphics[scale=1]{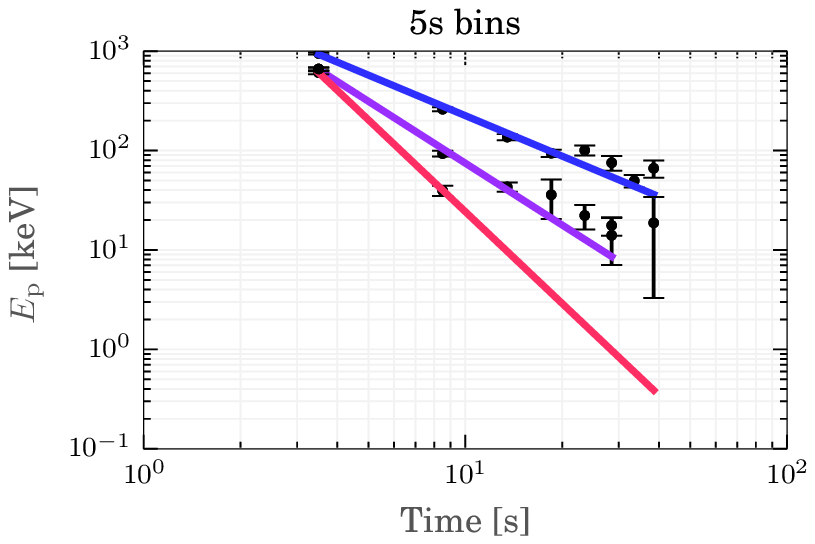}}\subfigure[]{\includegraphics[scale=1]{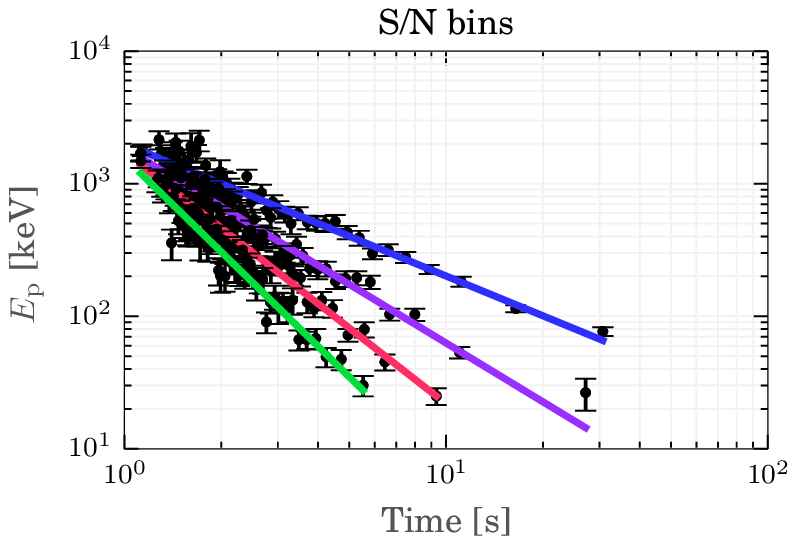}}
\subfigure[]{\includegraphics[scale=1]{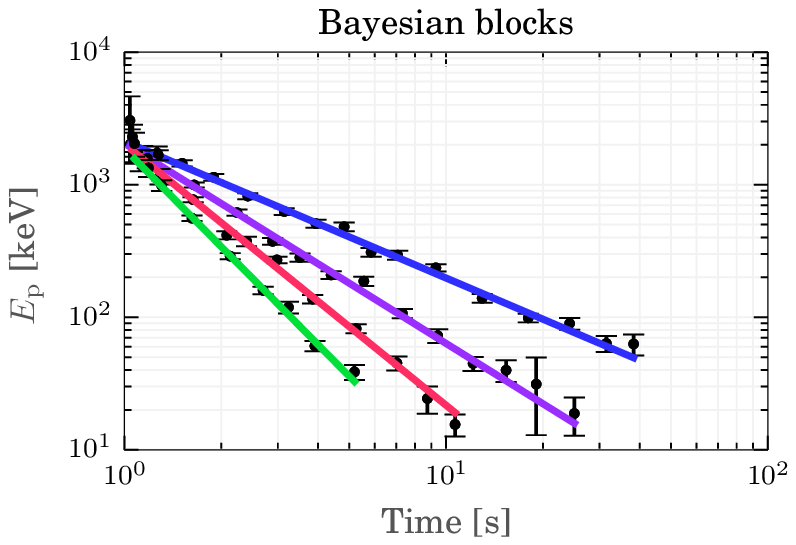}}\subfigure[]{\includegraphics[scale=1]{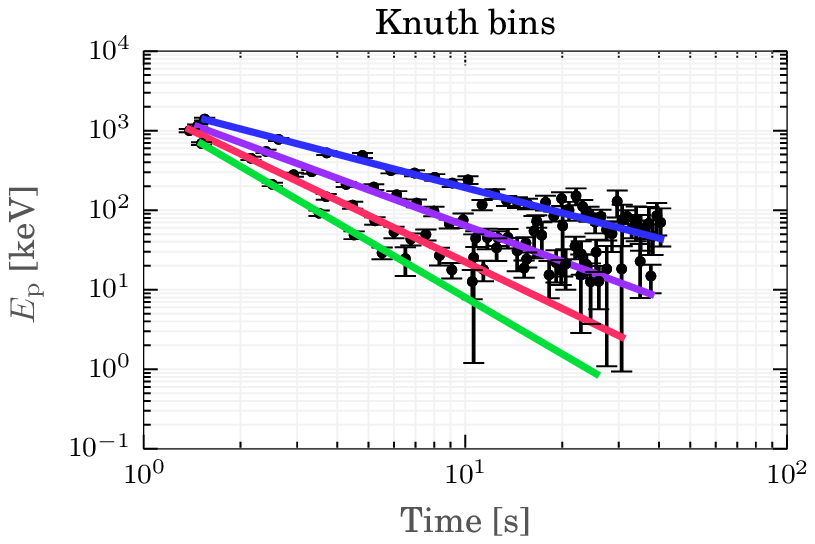}}
\caption{The reconstructed evolution of $\Ep$ for each of the binning methods with simulated $\alpha=0$. The various fit lines indicate the reconstructed evolution for $\gamma=$1 ({\it blue}), 1.5 ({\it purple}), 2 ({\it pink}), and 2.5 ({\it green}). Some data points are missing due to the fitting engine failing to converge.}
\label{fig:alph0EpEvo}
\end{figure*}

\begin{figure*}
\begin{centering}
  \includegraphics[scale=1]{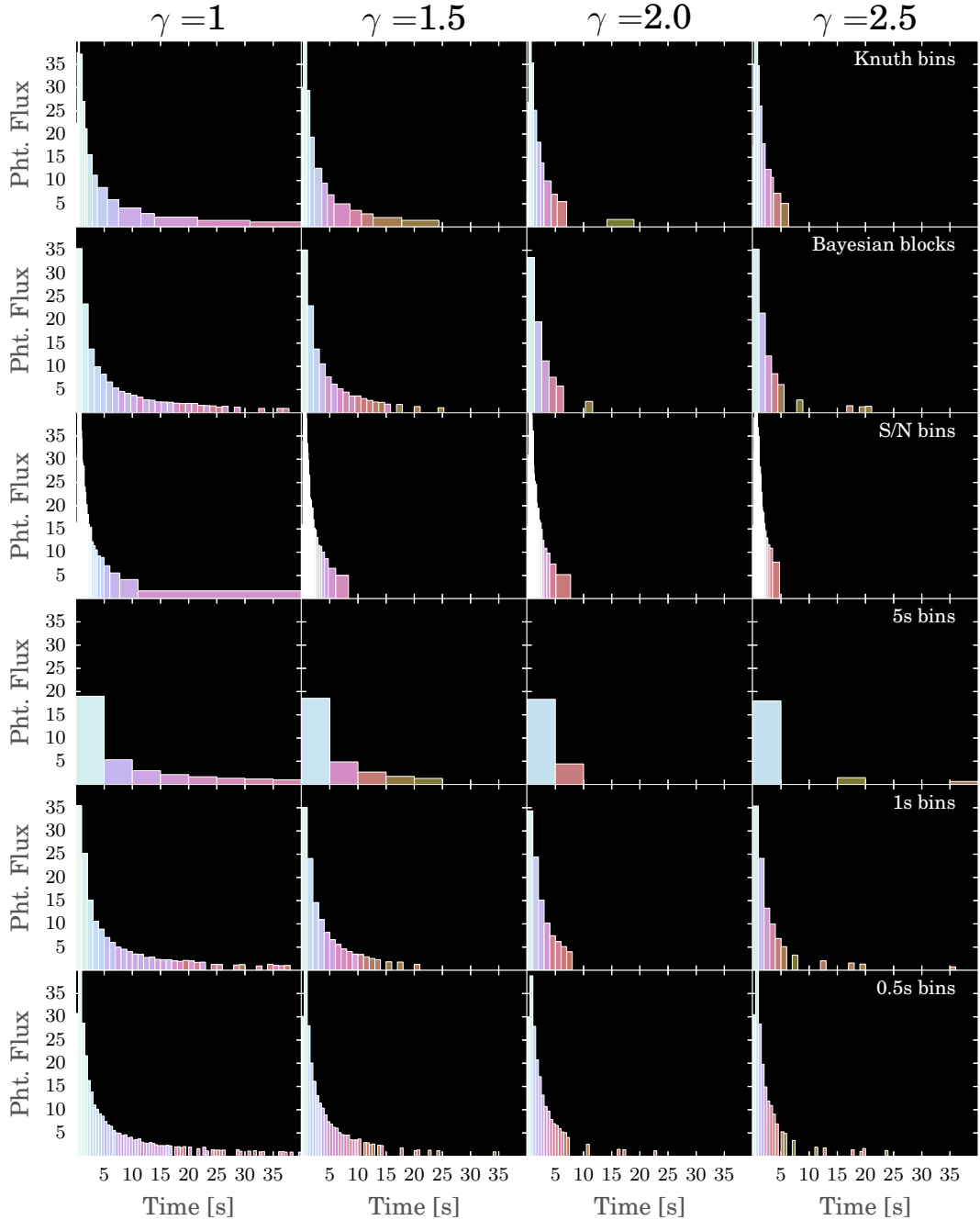}
\end{centering}
\caption{An example of the reconstructed flux and $\Ep$ from fits with
  $\alpha=-1$. Missing bins are due to the fitting engine failing to
  converge. It is obvious that coarse binnings have lower $\Ep$'s
  during the peak flux phase of the pulse. This is due to spectral
  averaging. When the evolution of $\Ep$ is fast ($\gamma=2,2.5$), it
  is difficult to fit spectral in the tail of the pulse because $\Ep$
  quickly moves out of the instrument's bandpass. However, evolution
  this fast is rarely observed in GRBs. These plots can be compared to
  Figure \ref{fig:exEpevo2} for the simulated $\Ep$ evolution.}

  \label{fig:epfit}
\end{figure*}

Using the fits for each set of simulations, the time evolution of
$\Ep$ is fit with a power law and the recovered temporal index is
compared with the simulated value. In general, the various binning
methods recover the simulated $\Ep$ evolution satisfactorily (see
Table \ref{tab:epEvo} and Figures \ref{fig:alph-1EpEvo},
\ref{fig:alph0EpEvo}, and \ref{fig:epfit}). There are, however noted
exceptions.

The bins produced by the S/N method systematically flatten the
evolution of $\Ep$. The origin of the effect is difficult to
ascertain. A subset of the simulated datasets were binned using a
signal-to-noise ratio of 100 to check how the ratio affected the
results. The reconstructed values of $\gamma$ were all flattened
compared to the simulated value. The fine time (CC$_{0.5,1}$) bins
both reconstructed the evolution well. On average, they steepened
$\gamma$ with some exceptions. The coarse (CC$_{5}$) bins
systematically steepened $\gamma$. Both the KB and BB methods are
accurate in reconstructing the evolution with no exceptions. This may
be due to the fact that they bin the data based on the inherent
temporal structure of the lightcurve. 

\begin{table}
\begin{tabular}{lccc}

 & &$\alpha=-1.$ & $\alpha=0.$ \\
\hline\hline
Binning & True  $\gamma$& Fitted $\gamma$  & Fitted $\gamma$  \\
\hline
0.5s bins &  \multirow{6}{*}{1.0}& $1.18\pm0.04$ &  $1.05\pm0.02$  \\
1s bins &  & $1.15\pm0.04$ &  $1.04\pm0.02$   \\
5s bins &  & $1.29\pm0.07$ & $1.35\pm0.03$  \\
Bayesian blocks &  & $1.10\pm0.04$   & $1.03\pm0.02$  \\
Knuth bins &  & $0.99\pm0.05$ &$1.06\pm0.02$  \\
S/N bins &  & $0.81\pm0.05$ & $1.00\pm0.02$  \\
\hline
0.5s bins & \multirow{6}{*}{1.5} & $1.56\pm0.05$ & $1.50\pm0.03$ \\
1s bins &  & $1.57\pm0.05$ & $1.50\pm0.03$  \\
5s bins &  & $2.10\pm0.09$ & $2.07\pm0.06$ \\
Bayesian blocks &  & $1.57\pm0.05$ & $1.51\pm0.03$ \\
Knuth bins &  & $1.54\pm0.05$ & $1.49\pm0.03$  \\
S/N bins &  & $1.22\pm0.08$ & $1.47\pm0.04$ \\
\hline
0.5s bins & \multirow{6}{*}{2.0} & $1.98\pm0.09$ & $1.98\pm0.04$ \\
1s bins &  & $2.11\pm0.10$  & $1.96\pm0.05$  \\
5s bins &  & $2.99\pm0.36$  & $3.07\pm0.14$  \\
Bayesian blocks &  & $2.12\pm0.11$ &  $1.97\pm0.05$  \\
Knuth bins &  & $1.95\pm0.10$ &$1.94\pm0.04$  \\
S/N bins &  & $1.73\pm0.10$ & $1.90\pm0.05$  \\
\hline
0.5s bins & \multirow{6}{*}{2.5} & $2.53\pm0.12$ &  $2.32\pm0.06$  \\
1s bins &  & $2.41\pm0.12$  & $2.35\pm0.07$   \\
5s bins &  & $2.10\pm0.16$   & $3.07\pm0.14$    \\
Bayesian blocks &  & $2.49\pm0.12$  & $2.46\pm0.07$   \\
Knuth bins &  & $2.47\pm0.14$  & $2.36\pm0.07$  \\
S/N bins &  & $2.20\pm0.16$ & $2.37\pm0.07$   \\
\end{tabular}
\caption{The reconstruction of $\Ep$ evolution in time for the tested binning methods.}
\label{tab:epEvo}
\end{table}

\subsection[]{Spectral Averaging}

Spectral averaging occurs when the evolution of the spectrum across
the duration of a time bin is summed. In the case of the Band
function, with its adaptable fit parameters, the spectrum resulting
from averaging its evolution across a time bin resembles a Band
function. This may not be the case for actual physical models and
therefore, it is pertinent to test how the spectral averaging of these
models appears when fit with a Band function. Moreover, if the
spectrum consist of multiple components that evolve in time
independently as has been shown
\citep{Guiriec:2011jr,Guiriec:2013hl,Axelsson:2012ic,Preece:2014ho,Burgess:2014db},
then fitting these time-averaged with the empirical Band function will
give no physical insight into the models at all. Investigating the
properties of physical model evolution is beyond the scope of this
work. The focus here is on how time-averaging of the Band function
affects the fitted values during data analysis.

\begin{figure*}
\subfigure[]{  \includegraphics[scale=1]{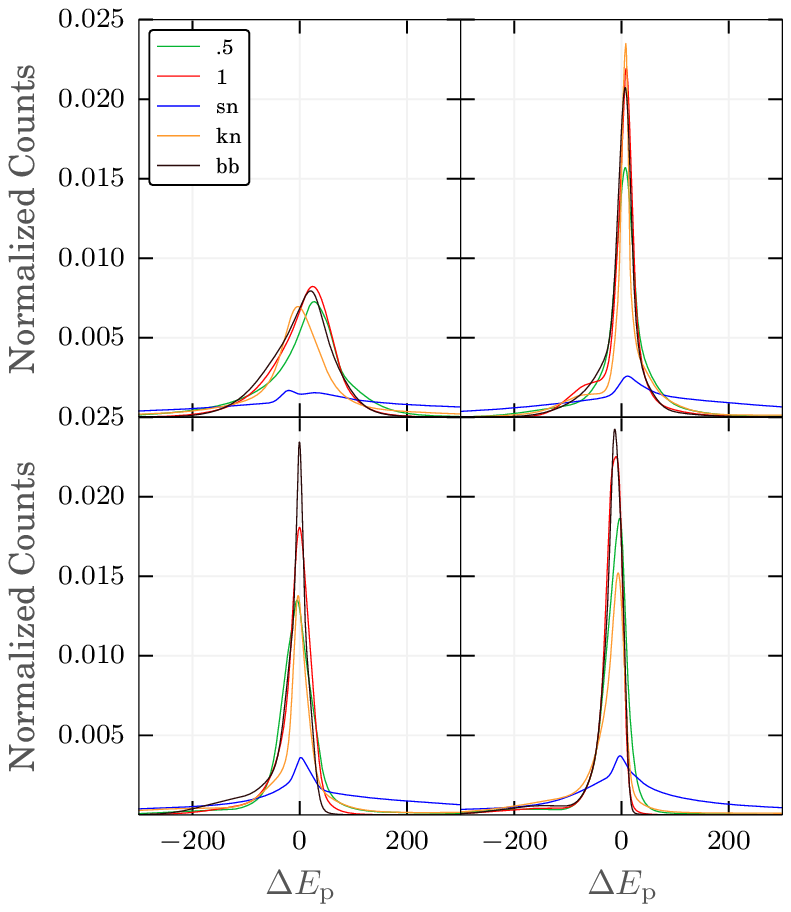}}\subfigure[]{
  \includegraphics[scale=1]{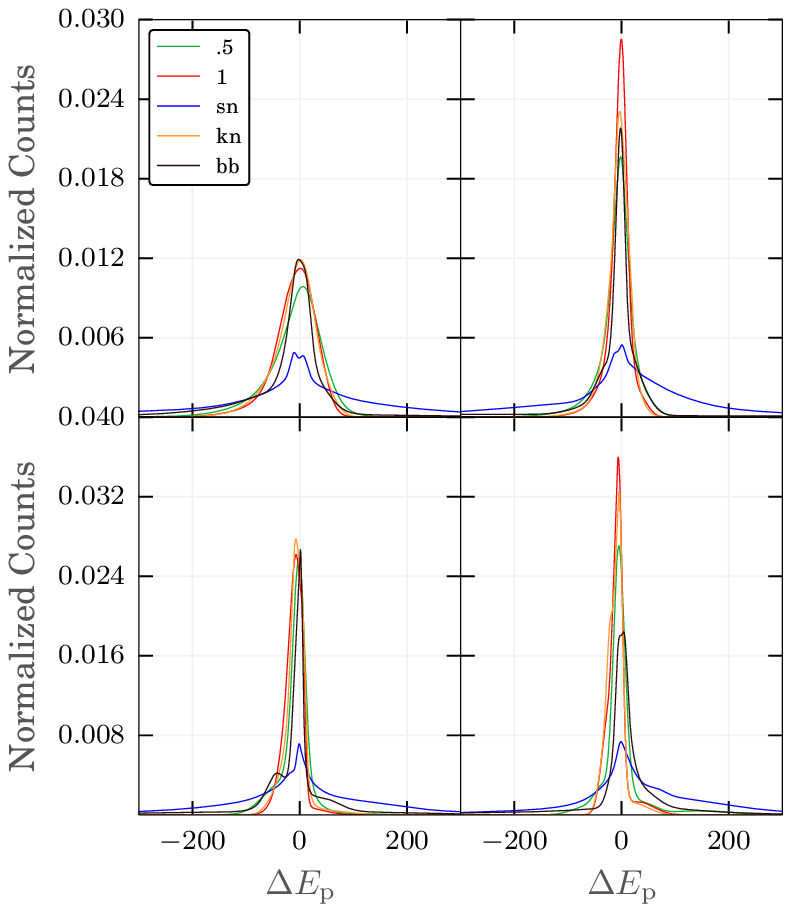}}
\caption{Continuous histograms of the shift of the fitted $\Ep$ from
  its simulated value. Clearly, the values obtained from S/N binning
  have a much larger spread than the other techniques. The fits made
  with Bayesian blocks have a better determination of $\Ep$ on
  average.}
  \label{fig:epshit0}
\end{figure*}

The result of spectral averaging is apparent in Figure \ref{fig:epfit}
where coarse time bins have a systematically lower $\Ep$ at the
beginning of the pulse. The fitted value of $\Ep$ is compared with its
simulated value in the center of the time bin in
Figure \ref{fig:epshit0}. The most evident feature is that around the peak
flux of the pulse, the S/N bins differ greatly from the simulated
value. The coarse CC bins also differ greatly a late times in the
pulse. While the overall evolution can be recovered in time, these
differences become important when calculating physical parameters from
the fit values.

\begin{figure*}
\subfigure[]{  \includegraphics[scale=1]{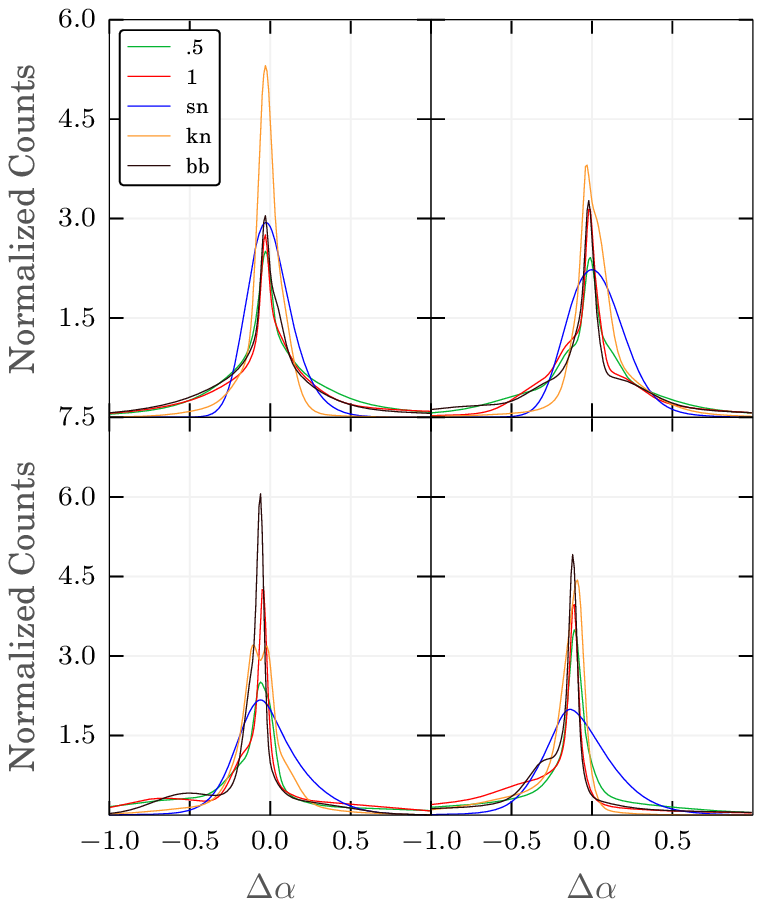}}\subfigure[]{
  \includegraphics[scale=1]{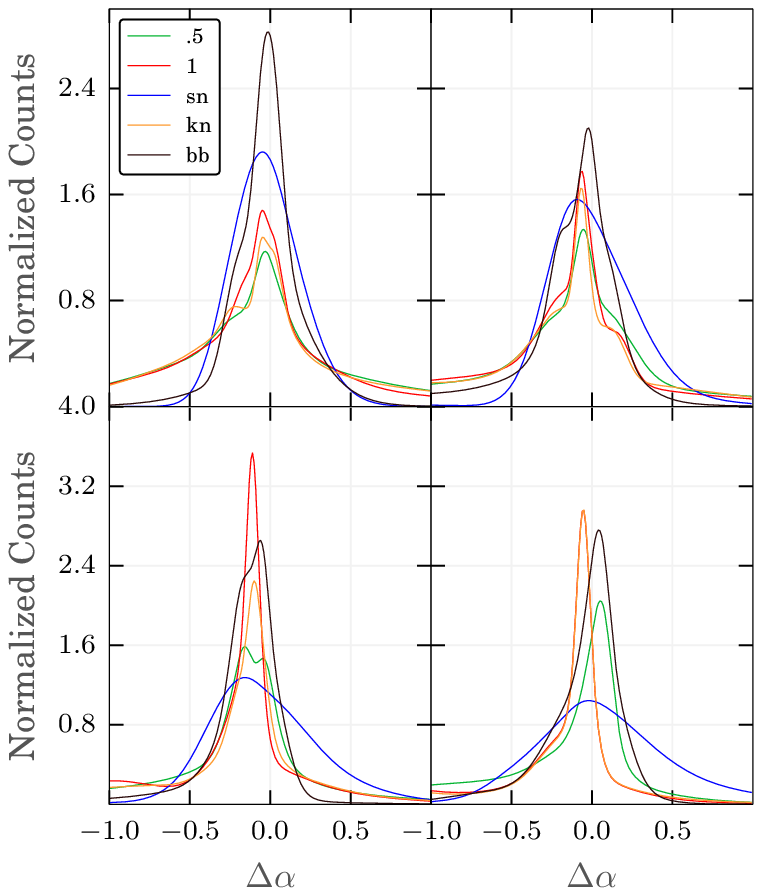}}
\caption{Continuous histograms of the shift of the fitted $\alpha$
  from its simulated value. As with $\Ep$ (see Figure
  \ref{fig:epshit0}), the bins made with S/N have a much broader
  distribution and are less accurate than the other methods at
  reconstructing the true value of $\alpha$.}
  \label{fig:ashit0}
\end{figure*}

Another interesting value to investigate is the Band function's
low-energy index, $\alpha$. This parameter is of interest because it
is often used to interpret the type of high-energy emission that is
occurring in the GRB jet \citep{Baring:2004wg,Preece:1998wk}. While
the value of $\alpha$ is held constant through out the simulated
pulse, spectral averaging can affect its value in the fit. As seen in
Figure \ref{fig:ashit0}, most binning methods reconstruct the value of
$\alpha$ accurately except near the tail of the pulse where the flux
is low. It is apparent that the S/N binning poorly reconstructs the
value around the peak similar to what is observed with
$\Ep$. Additionally, when the spectral evolution is very fast
($\gamma$=2,2.5), the recovered value of $\alpha$ systematically
shifted to the softer values regardless of the binning method. This
should be noted if it is found that $\Ep$ is evolving quickly in a
GRB.

\section{Conclusions}

This investigation of the temporal binning of GRB lightcurves for
spectral analysis has revealed several important factors to be
considered when choosing time bins for spectral analysis. Coarse
binning via the CC method of the data allows for a significant amount
of counts to be used in the fit. This does not; however, imply that
the fit is constrained or accurate. The differences of the fitted and
simulated values of $\Ep$ with coarse binning can lead to incorrectly
inferred physical values of the GRB jet. The overall trend of $\Ep$
can still be reconstructed with course binning but is unnecessary when
the finer CC bins are more accurate at reconstructing $\Ep$. Still,
the most accurate methods to bin the data are Knuth binning and
Bayesian blocks. These methods take different approaches to
determining the variability of the source but both are very accurate
in reconstructing the values and evolution of $\Ep$. Both methods
possess well prescribed reasoning for their form which can be used in
justifying the methods use in spectral analysis.

The only methods that have negative effects on analysis is that of S/N
bins and coarse CC bins. These bins poorly reconstruct both the
evolution and values of $\Ep$. With S/N bins, the method is also not
entirely objective. Bins of high flux are finer while bins of low flux
are wide. This may not reflect either the intrinsic variability of the
source or the underlying evolution of $\Ep$. Increasing the S/N ratio
for a subset of the simulated dataset only furthered the
problem. While it is possible that a value of S/N exists such that the
correct evolution can be reconstructed, with real data there is no way
to know what this correct value would be. It is entirely likely that
such a value would change from burst to burst where temporally varying
backgrounds and non-standard pulse shapes are common. Therefore, the
method of using S/N to bin data is cautioned against when there are
other methods that clearly give better results.

As mentioned above, none of these methods address a way to bin the
data based on spectral evolution. The fact that KB and BB bins can
accurately reconstruct the evolution of $\Ep$ and also have bins that
are based on the intrinsic source variability points to these methods
as an objective and accurate way to choose time bins. Both of these
qualities are essential in identifying the physical emission processes
and jet dynamics occurring in GRBs.  Moreover, the observation that
$\Ep$ is strongly correlated with energy flux in most GRBs further
motivates the choice of a binning method that can select bins based
upon the intrinsic flux evolution of the lightcurve. While the
evolution simulated here is simple, it is a common evolution observed
in GRBs. More complex evolutions may present problems for all methods.

Regarding the integrated properties of GRBs, based on this
investigation it is difficult to infer meaning to the integrated $\Ep$
of a GRB without knowing the proper emission mechanism of the
burst. This fact in combination with the studies of
\citet{Kocevski:2012kc,Nakar:2005ja,Band:2005vb} make using the
integrated properties of GRBs as cosmological tools difficult.  If the
proper emission mechanism is known, then the integrated spectra can be
calculated from first principles and integrated fits will be useful
for inferring physical properties. This is vitally important for
examining short and/or weak GRBs which do not allow for time-resolved
spectroscopy. Until this is accomplished, physical interpretations of
integrated spectra should be done with extreme
caution. 

\section*{Acknowledgments}
JMB would like to thank Valerie Connaughton, Michael Briggs, Felix
Ryde, and Rob Preece for useful discussions as well as the Oskar-Klein
Center for supporting this work. 

\bibliographystyle{mn2e}
\bibliography{bib}

\begin{thebibliography}{27}
\expandafter\ifx\csname natexlab\endcsname\relax\def\natexlab#1{#1}\fi

\bibitem[{Amati(2003)}]{Amati:2003wu}
Amati L., 2003, Chinese Journal of Astronomy and Astrophysics

\bibitem[{Arnaud(1996)}]{xspec}
Arnaud K., 1996, ASP Conf. Series, 101, 17

\bibitem[{Axelsson {et~al}\mbox{.}(2012)Axelsson, Baldini, Barbiellini, Baring,
  Bellazzini, Bregeon, Brigida, Bruel, Buehler, Caliandro, Cameron, Caraveo,
  Cecchi, Chaves, Chekhtman, Chiang, Claus, Conrad, Cutini,
  D{\textquoteright}Ammando, de~Palma, Dermer, do~Couto~e Silva, Drell,
  Favuzzi, Fegan, Ferrara, Focke, Fukazawa, Fusco, Gargano, Gasparrini,
  Gehrels, Germani, Giglietto, Giroletti, Godfrey, Guiriec, Hadasch, Hanabata,
  Hayashida, Hou, Iyyani, Jackson, Kocevski, Kuss, Larsson, Larsson, Longo,
  Loparco, Lundman, Mazziotta, McEnery, Mizuno, Monzani, Moretti, Morselli,
  Murgia, Nuss, Nymark, Ohno, Omodei, Pesce-Rollins, Piron, Pivato, Racusin,
  Rain{\`o}, Razzano, Razzaque, Reimer, Roth, Ryde, Sanchez, Sgr{\`o}, Siskind,
  Spandre, SPINELLI, Stamatikos, Tibaldo, Tinivella, Usher, Vandenbroucke,
  Vasileiou, Vianello, Vitale, Waite, Winer, Wood, Burgess, Bhat, Bissaldi,
  Briggs, Connaughton, Fishman, Fitzpatrick, Foley, Gruber, Kippen,
  Kouveliotou, Jenke, McBreen, McGlynn, Meegan, Paciesas, Pelassa, Preece,
  Tierney, Von~Kienlin, Wilson-Hodge, Xiong, \& Pe'Er}]{Axelsson:2012ic}
Axelsson M. {et~al.}, 2012, ApJL, 757, L31

\bibitem[{Band {et~al}\mbox{.}(1993)Band, Matteson, Ford, \&
  Schaefer}]{Band:1993wc}
Band D., Matteson J., Ford L., Schaefer B., 1993, ApJ

\bibitem[{Band \& Preece(2005)}]{Band:2005vb}
Band D., Preece R., 2005, ApJ

\bibitem[{Band(1997)}]{Band:1997}
Band D.~L., 1997, ApJ, 486, 928

\bibitem[{Baring \& Braby(2004)}]{Baring:2004wg}
Baring M.~G., Braby M., 2004, ApJ

\bibitem[{Burgess {et~al}\mbox{.}(2014)Burgess, Preece, Connaughton, Briggs,
  Goldstein, Bhat, Greiner, Gruber, Kienlin, Kouveliotou, McGlynn, Meegan,
  Paciesas, Rau, Xiong, Axelsson, Baring, Dermer, Iyyani, Kocevski, Omodei,
  Ryde, \& Vianello}]{Burgess:2014db}
Burgess J.~M. {et~al.}, 2014, ApJ, 784, 17

\bibitem[{Crider {et~al}\mbox{.}(1998)Crider, Liang, Preece, Briggs, Pendleton,
  \& Band}]{Crider:1998uf}
Crider A., Liang E., Preece R., Briggs M., Pendleton G., Band D., 1998, APS
  April Meeting Abstracts, -1

\bibitem[{Ghirlanda, Ghisellini \& Firmani(2005)Ghirlanda, Ghisellini, \&
  Firmani}]{Ghirlanda:2005uy}
Ghirlanda G., Ghisellini G., Firmani C., 2005, MNRAS

\bibitem[{Ghirlanda, Ghisellini \& Lazzati(2004)Ghirlanda, Ghisellini, \&
  Lazzati}]{Ghirlanda:2004tq}
Ghirlanda G., Ghisellini G., Lazzati D., 2004, ApJL

\bibitem[{Goldstein {et~al}\mbox{.}(2012)Goldstein, Goldstein, Burgess,
  Burgess, Preece, Preece, Briggs, Guiriec, van~der Horst, Connaughton,
  Wilson-Hodge, Paciesas, Meegan, von Kienlin, Bhat, Bissaldi, Chaplin, Diehl,
  Fishman, Fitzpatrick, Foley, Gibby, Giles, Greiner, Gruber, Kippen,
  Kouveliotou, McBreen, McGlynn, Rau, \& Tierney}]{Goldstein:2012go}
Goldstein A. {et~al.}, 2012, ApJS, 199, 19

\bibitem[{Golenetskii {et~al}\mbox{.}(1983)Golenetskii, Mazets, Aptekar, \&
  Ilinskii}]{1983Natur.306..451G}
Golenetskii S.~V., Mazets E.~P., Aptekar R.~L., Ilinskii V.~N., 1983, Nature,
  306, 451

\bibitem[{Guiriec {et~al}\mbox{.}(2010)Guiriec, Briggs, Connaugthon, Kara,
  Daigne, Kouveliotou, van~der Horst, Paciesas, Meegan, Bhat, Foley, Bissaldi,
  Burgess, Chaplin, Diehl, Fishman, Gibby, Giles, Goldstein, Greiner, Gruber,
  von Kienlin, Kippen, McBreen, Preece, Rau, Tierney, \&
  Wilson-Hodge}]{2010ApJ...725..225G}
Guiriec S. {et~al.}, 2010, ApJ, 725, 225

\bibitem[{Guiriec {et~al}\mbox{.}(2011)Guiriec, Connaughton, Briggs, Burgess,
  Ryde, Daigne, Meszaros, Goldstein, McEnery, Omodei, Bhat, Bissaldi,
  Camero-Arranz, Chaplin, Diehl, Fishman, Foley, Gibby, Giles, Greiner, Gruber,
  von Kienlin, Kippen, Kouveliotou, McBreen, Meegan, Paciesas, Preece, Rau,
  Tierney, van~der Horst, \& Wilson-Hodge}]{Guiriec:2011jr}
Guiriec S. {et~al.}, 2011, ApJL, 727, L33

\bibitem[{Guiriec {et~al}\mbox{.}(2013)Guiriec, Daigne, Hascoet, Vianello,
  Ryde, Mochkovitch, Kouveliotou, Xiong, Bhat, Foley, Gruber, Burgess, McGlynn,
  McEnery, \& Gehrels}]{Guiriec:2013hl}
Guiriec S. {et~al.}, 2013, ApJ, 770, 32

\bibitem[{Knuth(2006)}]{2006physics...5197K}
Knuth K.~H., 2006, arXiv.org, 5197

\bibitem[{Kocevski(2012)}]{Kocevski:2012kc}
Kocevski D., 2012, ApJ, 747, 146

\bibitem[{Kocevski, Ryde \& Liang(2003)Kocevski, Ryde, \&
  Liang}]{2003ApJ...596..389K}
Kocevski D., Ryde F., Liang E., 2003, ApJ, 596, 389

\bibitem[{Liang \& Kargatis(1996)}]{Liang:1996cl}
Liang E., Kargatis V., 1996, Nature, 381, 49

\bibitem[{Meegan {et~al}\mbox{.}(2009)Meegan, Lichti, Bhat, Bissaldi, Briggs,
  Connaughton, Diehl, Fishman, Greiner, Hoover, van~der Horst, von Kienlin,
  Kippen, Kouveliotou, McBreen, Paciesas, Preece, Steinle, Wallace, Wilson, \&
  Wilson-Hodge}]{2009ApJ...702..791M}
Meegan C. {et~al.}, 2009, The Astrophysical Journal, 702, 791

\bibitem[{Nakar \& Piran(2005)}]{Nakar:2005ja}
Nakar E., Piran T., 2005, MNRASL, 360, L73

\bibitem[{Norris {et~al}\mbox{.}(1986)Norris, Share, Messina, Dennis, Desai,
  Cline, Matz, \& Chupp}]{1986ApJ...301..213N}
Norris J.~P., Share G.~H., Messina D.~C., Dennis B.~R., Desai U.~D., Cline
  T.~L., Matz S.~M., Chupp E.~L., 1986, ApJ, 301, 213

\bibitem[{Preece {et~al}\mbox{.}(1998)Preece, Briggs, Mallozzi, \&
  Pendleton}]{Preece:1998wk}
Preece R., Briggs M.~S., Mallozzi R., Pendleton G.~N., 1998, ApJL

\bibitem[{Preece {et~al}\mbox{.}(2014)Preece, Burgess, Von~Kienlin, Bhat,
  Briggs, Byrne, Chaplin, Cleveland, Collazzi, Connaughton, Diekmann,
  Fitzpatrick, Foley, Gibby, Giles, Goldstein, Greiner, Gruber, Jenke, Kippen,
  Kouveliotou, McBreen, Meegan, Paciesas, Pelassa, Tierney, van~der Horst,
  Wilson-Hodge, Xiong, Younes, Yu, Ackermann, Ajello, Axelsson, Baldini,
  Barbiellini, Baring, Bastieri, Bellazzini, Bissaldi, Bonamente, Bregeon,
  Brigida, Bruel, Buehler, Buson, Caliandro, Cameron, Caraveo, Cecchi, Charles,
  Chekhtman, Chiang, Chiaro, Ciprini, Claus, Cohen-Tanugi, Cominsky, Conrad,
  D{\textquoteright}Ammando, de~Angelis, de~Palma, Dermer, Desiante, Digel,
  Di~Venere, Drell, Drlica-Wagner, Favuzzi, Franckowiak, Fukazawa, Fusco,
  Gargano, Gehrels, Germani, Giglietto, Giordano, Giroletti, Godfrey, Granot,
  Grenier, Guiriec, Hadasch, Hanabata, Harding, Hayashida, Iyyani, Jogler,
  J{\'o}hannesson, Kawano, Kn{\"o}dlseder, Kocevski, Kuss, Lande, Larsson,
  Larsson, Latronico, Longo, Loparco, Lovellette, Lubrano, Mayer, Mazziotta,
  Michelson, Mizuno, Monzani, Moretti, Morselli, Murgia, Nemmen, Nuss, Nymark,
  Ohno, Ohsugi, Okumura, Omodei, Orienti, Paneque, Perkins, Pesce-Rollins,
  Piron, Pivato, Porter, Racusin, Rain{\`o}, Rando, Razzano, Razzaque, Reimer,
  Reimer, Ritz, Roth, Ryde, Sartori, Scargle, Schulz, Sgr{\`o}, Siskind,
  Spandre, SPINELLI, Suson, Tajima, Takahashi, Thayer, Thayer, Tibaldo,
  Tinivella, Torres, Tosti, Troja, Usher, Vandenbroucke, Vasileiou, Vianello,
  Vitale, Werner, Winer, Wood, \& Zhu}]{Preece:2014ho}
Preece R. {et~al.}, 2014, Science, 343, 51

\bibitem[{Scargle {et~al}\mbox{.}(2013)Scargle, Norris, Jackson, \&
  Chiang}]{2013ApJ...764..167S}
Scargle J.~D., Norris J.~P., Jackson B., Chiang J., 2013, ApJ, 764, 167

\bibitem[{VanderPlas {et~al}\mbox{.}(2012)VanderPlas, Connolly, Ivezic, \&
  Gray}]{VanderPlas:gg}
VanderPlas J., Connolly A.~J., Ivezic Z., Gray A., 2012, 2012 Conference on
  Intelligent Data Understanding (CIDU), 47

\end{thebibliography}


\newpage


\label{lastpage}

\end{document}